\documentclass[twocolumn, tighten, dvipsnames, floatfix]{aastex63}

\usepackage[T1]{fontenc}
\usepackage[]{xcolor}
\usepackage{booktabs}
\usepackage{physics}
\usepackage{wasysym}
\usepackage{longtable}
\usepackage{threeparttablex} 
\usepackage{tabularx}
\usepackage{graphicx}

\usepackage{array}
\usepackage{amsmath}
\usepackage{changepage}
\usepackage{lineno, ulem} 

\usepackage[flushmargin]{footmisc}
\usepackage{afterpage}
\usepackage{subfigure}
\usepackage{multirow}

\DeclareUnicodeCharacter{FF0C}{,}

\newcommand{\kms}{\mbox{$\>{\rm km\, s^{-1}}$}}

\def\arcsec{\hbox{$^{\hbox{\rlap{\hbox{\lower4pt\hbox{$\,\prime\prime$}}
          }}}$} \ }

\def\arcmin{\hbox{$^{\hbox{\rlap{\hbox{\lower4pt\hbox{$\;\prime$}}
          }\hbox{$\frown$}}}$}}

\usepackage{lipsum}
\usepackage{float}
\usepackage[misc]{ifsym}

\shorttitle{The Low-$\alpha$ Splash}
\shortauthors{Borbolato et al.}

\begin{document}

\title{The Low-$\alpha$ Splash Population in the Milky Way}

\correspondingauthor{Lais Borbolato (laisborbolato@usp.br), Jo\~ao A. S. Amarante (joaoant@gmail.com) }

\author[0000-0003-3382-1051]{Lais Borbolato}
\affil{Universidade de S\~ao Paulo, Instituto de Astronomia, Geof\'isica e Ci\^encias Atmosf\'ericas, Departamento de Astronomia, \\ SP 05508-090, S\~ao Paulo, Brasil}

\author[0000-0002-7662-5475]{Jo\~ao A. S. Amarante}
\affil{Department of Astronomy, School of Physics and Astronomy, \\ Shanghai Jiao Tong University, 800 Dongchuan Road, Shanghai, 200240, China}
\affil{State Key Laboratory of Dark Matter Physics, School of Physics and Astronomy，\\ Shanghai Jiao Tong University, Shanghai, 200240, China}

\author[0000-0002-0537-4146]{H\'elio D. Perottoni}
\affil{Observat\'{o}rio Nacional, MCTI, Rua Gal. Jos\'{e} Cristino 77, Rio de Janeiro, 20921-400, RJ, Brazil}

\author[0000-0001-7479-5756]{Silvia Rossi}
\affil{Universidade de S\~ao Paulo, Instituto de Astronomia, Geof\'isica e Ci\^encias Atmosf\'ericas, Departamento de Astronomia, \\ SP 05508-090, S\~ao Paulo, Brasil}

\author[0000-0001-7902-0116]{Victor P. Debattista}
\affil{Jeremiah Horrocks Institute, University of Lancashire, Preston, PR1 2HE, UK}

\author[0000-0001-5017-7021]{Zhao-Yu Li}
\affil{Department of Astronomy, School of Physics and Astronomy, \\ Shanghai Jiao Tong University, 800 Dongchuan Road, Shanghai, 200240, China}
\affil{State Key Laboratory of Dark Matter Physics, School of Physics and Astronomy，\\ Shanghai Jiao Tong University, Shanghai, 200240, China}

\author[0000-0003-3523-7633]{Nathan Deg}
\affil{Department of Physics, Engineering Physics, and Astronomy, Queen’s University, Kingston ON K7L 3N6, Canada}

\author[0000-0002-3343-6615]{Tigran Khachaturyants}
\affil{Department of Astronomy, School of Physics and Astronomy, \\ Shanghai Jiao Tong University, 800 Dongchuan Road, Shanghai, 200240, China}
\affil{State Key Laboratory of Dark Matter Physics, School of Physics and Astronomy，\\ Shanghai Jiao Tong University, Shanghai, 200240, China}


\begin{abstract}

The Milky Way metal-rich halo-like component, also known as the Splash, consists of old (Age $>$ 10 Gyr), metal-rich ([Fe/H] $> -0.7$), high-$\alpha$ stars, i.e. thick disk-like chemistry, on halo-like orbits (eccentricity > 0.6). Its origin is linked to stars formed in the disk and dynamically heated by either internal or external agents. In this work, we investigate its low-$\alpha$ counterpart, the \textit{low-$\alpha$ Splash}, motivated by recent findings of an old thin disk population. We conjecture that any mechanism capable of heating disk stars should affect both present-day high- and low-$\alpha$ old populations. Using data from the APOGEE DR17 spectroscopic catalog, we identify metal-rich low-$\alpha$ stars with halo-like kinematics similar to those of the classical high-$\alpha$ Splash. We investigate their possible heating mechanisms using GASTRO suite of simulations, which allows us to explore the effects of star-forming clumps as well as a major merger in the proto-disk of a Milky Way analog galaxy. Our main results show that only clumpy Milky Way models are able to produce Splash populations, 
including the low-$\alpha$ counterpart, whereas the model including only the merger and without an early clumpy phase fails to produce these populations. In the models, the low-$\alpha$ Splash corresponds to a subset of the old thin disk that was dynamically heated by the same mechanism responsible for the formation of the high-$\alpha$ Splash.
 
\end{abstract}



\section{Introduction}
\label{sec:intro}

The mechanisms driving the formation and evolution of the Milky Way (MW) over the past 13 billion years continue to be extensively investigated in the field of Galactic Archaeology (e.g., \citealt{Freeman2002, BlandHawthorn2016, Barbuy2018, Helmi2020}). A key approach involves studying the chemodynamical properties of the halo, bulge, and disk, focusing on populations that retain signatures of major evolutionary events, such as the dynamically heated disk population. Initially identified as metal-rich halo stars ([Fe/H] $> -1.0$; \citealt{Sheffield2012, Hawkins2015}), these objects were interpreted as stars ejected from the Galactic disk. The advent of the \textit{Gaia} era \citep{GaiaMission} and large spectroscopic surveys enabled the identification of hundreds of similar stars, and characterized these as an old, high-$\alpha$ population (e.g., \citealt{Bonaca2017, DiMatteo2019, Gallart2019, Belokurov2020, Bonaca2020}), consistent with a dynamically heated early disk.

Several works have sought to understand the formation of this population, leading to the prevailing interpretation that these stars were born in the early MW disk and then dynamically heated by the last major merger with the dwarf galaxy Gaia-Sausage/Enceladus (GSE; \citealt{belokurov2018, Haywood2018, helmi2018}), 9–11 Gyr ago (\citealt{Gallart2019, Montalban2021}). This population was dubbed ``Splash'' by \citet{Belokurov2020}, in reference to its likely formation pathway through a massive merger \citep[e.g][]{Grand2020,Liao2024, orkney+2025}. Contrary to this main interpretation, several mechanisms have been suggested for the formation of the Splash: i) scattering by massive star-forming clumps in early disk galaxies without the need for a merger event \citep{Amarante2020a}; ii) scattering by minor mergers (\citealt{Dillamore2022, Kisku2025}), and iii) an origin tied to the dispersion-dominated nature of the proto-galaxy (\citealt{Buder2025}).


Historically, the high-$\alpha$/thick\footnote{Here, we use low-$\alpha$ (high-$\alpha$) disk as a synonym for the chemical thin (thick) disk.} and low-$\alpha$/thin disks have been interpreted as a sequential formation process (e.g., \citealt{Chiappini1997, Grisoni2017}), where the thick disk formed first and the thin disk later, after dilution of the star-forming gas by accretion events such as major mergers \citep[e.g.,][]{Spitoni2019} or cosmic filaments \citep{Agertz2021, Renaud2021}. However, recent observations suggest that the early disk was not composed solely of a high-$\alpha$ population, but that low-$\alpha$ stars co-formed with the high-$\alpha$ disk during the first billion years of the MW (\citealt{Aguirre2018, Beraldo2021, Nepal2024, Gent2024, Borbolato2025}). \citet{Borbolato2025} showed that low-$\alpha$ stars predate the GSE merger, which leads us to conjecture that the same dynamical heating mechanism responsible for the high-$\alpha$ Splash should also have impacted the old low-$\alpha$ disk, potentially forming a low-$\alpha$ counterpart to the Splash. While recent studies report low-$\alpha$ stars on halo-like orbits (\citealt{Ortigoza2023, Nepal2024, Filion2025}), their origin remains debated, and its direct association with the high-$\alpha$ Splash population has not yet been established.


Until now, two models have been proposed to explain the formation of the ancient thin disk component ($z \sim 2$). One is the revised parallel model presented by \citet{Grisoni2026}, while the other is the clumpy model \citep{Beraldo2020, amarante+2026}. The latter also reproduces the chemo-geometric thin and thick disks \citep{Clarke2019, Beraldo2020}, and the chemistry of the bulge \citep{Debattista2023}. The possibility that the MW may have hosted clumps is also highly motivated by observations of clumpy galaxies (e.g., \citealt{Bournaud2008, Puech2010, Guo2015, Zanella2019, Adams2025, Sok2025}), with a peak around redshift z $\sim$ 2 (e.g., \citealt{Shibuya2016, Sattari2023, delaVega2025}), and the identification of galaxies analogous to MW progenitors hosting clumps \citep{Tan2025}. Also, \citet{daSilva2025} report that one very metal-poor r-process enriched star may have formed in a clump present in the early Galaxy. The clumpy model has proven to be a mechanism that likely played an important role in the formation of the MW disk and should therefore be considered in our investigation.

In this Letter, we demonstrate the existence of a low-$\alpha$ Splash population. We compare it with the canonical high-$\alpha$ Splash and explore possible formation pathways using the Gaia–EncelAdus–Sausage Timing, chemistRy and Orbit (GASTRO; \citealt{Amarante2022, amarante+2026}) simulations, examining the effects of star-forming clumps and a GSE-like merger on a MW-like disk. This Letter is organized as follows: Section \ref{sec:data} presents the observational and simulated data. Section \ref{sec:results} presents the results found, and they are discussed in Section \ref{sec:discussion}. Finally, Section \ref{conclusion} summarizes our main conclusions.

\begin{figure*}[ht!]
    \centering
    \includegraphics[width=2.1\columnwidth]{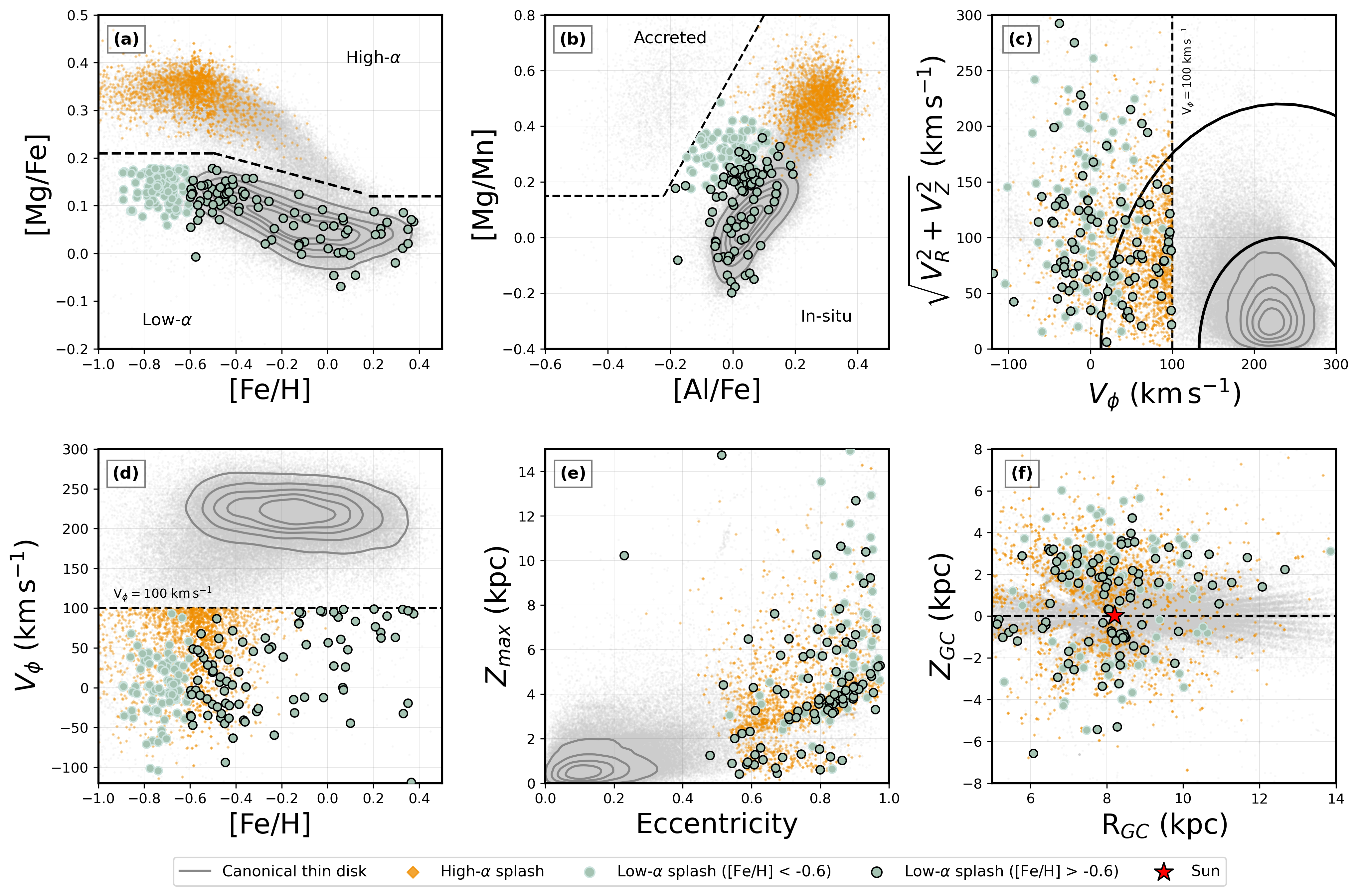}
    \caption{Projection of the Low-$\alpha$ Splash (green dots) and High-$\alpha$ Splash (orange points) in the \textbf{(a)} [Mg/Fe]--[Fe/H], where the dotted line indicates the separation between high- and low-$\alpha$ regions; \textbf{(b)} [Mg/Mn]--[Al/Fe], where the dashed line represent the limit between accreted and in-situ regions; \textbf{(c)} Toomre diagram, $\sqrt{V_R^2 + V_Z^2} \text{ versus } V_{\phi}$. The vertical dashed line is $V_{\phi}$ = 100\,km\,s$^{-1}$. The innermost curve represents $\rVert v - v_{circ} \rVert = 130$\,km\,s$^{-1}$ and the outermost curve is $\rVert v - v_{circ} \rVert = 220$\,km\,s$^{-1}$; \textbf{(d)} $V_{\phi}$--[Fe/H] space, where the horizontal dashed line is $V_{\phi}$ = 100\,km\,s$^{-1}$; \textbf{(e)} $Z_{\text{max}}$--Eccentricity and \textbf{(f)} $Z_{\text{GC}}$--$R_{\text{GC}}$, where the horizontal dashed line represents the Galactic plane ($Z_{\text{GC}}$ = 0\,kpc), and the red star indicates the position of the Sun. In all panels, the gray points are the APOGEE sample, and the gray contours indicate the density of the low-$\alpha$ region. The difference between light green points and green points with a black border is only the range of metallicity of the stars.}
    \label{Figure1_nova}
\end{figure*}

\section{Data}
\label{sec:data}

\subsection{Observational data}
\label{subsec:observationaldata}

We use data from the Apache Point Observatory Galactic Evolution Experiment (APOGEE; \citealt{apogee2017}) Data Release 17 (DR17; \citealt{APOGEEdr17}) spectroscopic survey, utilizing its high resolution ($R \sim 22,000$) near-infrared spectrograph \citep{Wilson2019} and $H$-band (1.51$\mu$m--1.69$\mu$m) coverage to distinguish the MW disk's high- and low-$\alpha$ populations \citep{imig+2023}.

Astrometric parameters are obtained from \textit{Gaia DR3} \citep{GAIADR32023}, with distances derived via the spectro-photometric \texttt{StarHorse} Bayesian code \citep{Queiroz2023}. Orbital integrations are performed for 20\,Gyr forward using the \texttt{AGAMA} library \citep{Vasiliev2019}, adopting the Galactic potential model of \citet{mcmillan2017}. We assume the Sun's position with respect to the Galactic Center to be $(X, Y, Z)_{\rm GC} = (-8.2, 0.0, 0.0)$ kpc \citep{BlandHawthorn2016}, with the local circular velocity $v_{circ} = 232.8$\kms \citep{mcmillan2017}, and the solar motion relative to the local standard of rest of $(U_{\odot}, V_{\odot}, W_{\odot}) = (11.10, 12.24, 7.25)$\kms \citep{schonrich2010}. Here, we consider that positive azimuthal velocity ($V_{\phi} > 0$\,\kms) corresponds to the sense of rotation of the disk.

Our sample consists of red giants ($4000 < T_\text{eff} \text{(K)} < 6500$, $\log g < 3$) with a good signal-to-noise ratio (\textit{S/N} $>$ 50) and sources with no spectral or parameter issues (\texttt{STARFLAG} == 0, \citealt{jonsson2020}). We require unflagged abundances for [Fe/H], [Mg/Fe], [Mg/Mn], and [Al/Fe] (i.e., flagged == 0). We adopt the recommended cut on the renormalized unit weight error ($\texttt{RUWE} \leq 1.4$; \citealt{Lindegren2020a}), a fractional distance uncertainty of $d/\sigma_{d} > 5$ and \texttt{parallax\_over\_error} $> 5$. After applying these quality cuts, the APOGEE sample contains 108,911 stars 


\subsection{MW-analogues with GASTRO Library}
\label{subsec:GASTROdata}

We use a subset of the GASTRO (\citealt{Amarante2022, amarante+2026}) library suite of simulations to explore the formation of the low$-\alpha$ Splash population. GASTRO consists of N-body + smoothed particle hydrodynamics (SPH) simulations designed to investigate the impact of a single merger event aimed at reproducing a GSE analog, as well as the role of high-star-formation clumps on the structure and evolution of an MW-like galaxy. We refer to \citet{Amarante2022} for a detailed description of the subgrid physics and initial conditions setup.

Here, we consider models with and without a merger. For both cases, we also vary the supernova feedback\footnote{Either 20\% or 80\% of the $10^{51}$ erg injected per supernova, the former allowing clump formation.}, which regulates the clump formation in the first Gyr. This results in four models: \texttt{Isolated nonClumpy}, \texttt{Isolated Clumpy}, \texttt{nonClumpy+merger}, \texttt{Clumpy+merger}, and they were introduced in \citet{amarante+2026} as {\it nc.iso}, {\it c.iso}, {\it nc.r.c03}, and {\it c.r.c03}, respectively. Below, we describe the main properties of these models (see \citealt{amarante+2026} for more details).\par

In the merger models, the satellite's orbit circularity\footnote{Defined as $\eta=L_z/L_c(E)$, where $L_z$ is the satellite's initial angular momentum, and $L_c(E)$ is the angular momentum of a planar circular orbit of energy $E$.} is set to $\eta=0.3$, and has a total dark matter and initial gas mass of $8.83\times10^{10}{\rm M_{\odot}}$ and $1.4\times10^9  {\rm M_{\odot}}$, respectively. The orbital evolution of the satellite is identical in both models; there are two clear apocentric passages, with the first pericenter being at $t\approx1.6$ Gyr, and the satellite is fully disrupted at $t\approx 3.2$ Gyr. At the first pericentric passage, the stellar mass of the satellite in the \texttt{nonClumpy+merger} and \texttt{Clumpy+merger} models is $3.15\times10^8{\rm M_{\odot}}$ and $8.97\times10^8{\rm M_{\odot}}$, respectively. These stellar masses are well within the estimated GSE stellar mass (see \citealt{Limberg2022} for a discussion). \par

The star-forming clumps in the models \texttt{Isolated Clumpy} and \texttt{Clumpy+merger} have a typical mass ranging from $0.3-1\times10^8{\rm M_{\odot}}$, and they typically survive for $\approx35-300$ Myr \citep{garver+2023}, in agreement with values measured from clumpy galaxies observed at redshift $z \approx 1-2$ \citep[e.g.][]{claeyssens+2023}. The clumps contribute up to 10-20\% of the total stellar mass of the disk in the first 0.5\,Gyr, and afterwards, they contribute less than 5\% and cease to exist at $\approx 3$\,Gyr (see Figure 1 of \citealt{amarante+2026}). All the models are run for 10\,Gyr, and we use the last snapshot to represent the ``present day''. All models show a clear chemical $\alpha$-bimodality in the disk, except for the \texttt{Isolated nonClumpy} model \citep{amarante+2026}. 


\section{Results}
\label{sec:results} 

\subsection{The observed Low-$\alpha$ Splash population}

To identify the low-$\alpha$ Splash members in APOGEE data, we select low- and high-$\alpha$ disks using [$\mathrm{Mg/Fe}$]--[$\mathrm{Fe/H}$] (Figure~\ref{Figure1_nova}, top-left) for stars with [Fe/H] $> -1.0$. Stars above (below) the black dashed line are classified as high-$\alpha$ (low-$\alpha$). We remove stars in the upper-left region of [$\mathrm{Mg/Mn}$]--[$\mathrm{Al/Fe}$] plane, associated with unevolved or accreted populations (\citealt{Hawkins2015b, Horta2021}), exclude cluster/dwarf galaxy members (\texttt{MEMBERFLAG} == 0), and restrict to $R_{\mathrm{GC}} > 5$\,kpc to avoid inner Galactic regions, resulting in 68,734 low-$\alpha$ and 25,676 high-$\alpha$ giant stars.

\begin{figure}
    \centering
    \includegraphics[width=\columnwidth]{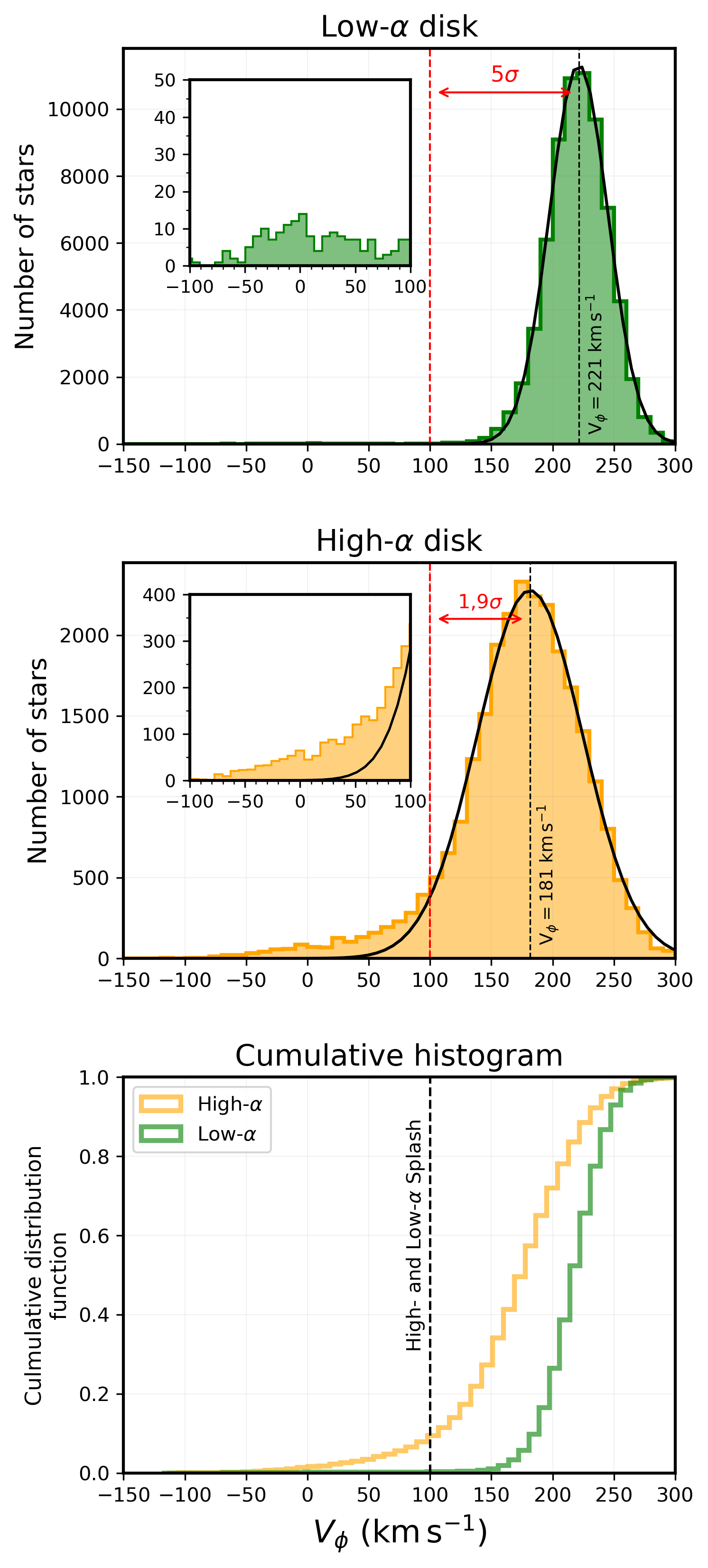}
    \caption{Distribution of azimuthal velocity $V_{\phi}$ for the low- and high-$\alpha$ samples in the upper and middle panels, respectively. The inner panels show a zoom on the left tail region of each distribution. The bottom panel shows the cumulative distribution of $V_{\phi}$ for both samples. The dashed vertical line at $V_{\phi} = 100$\,km\,s$^{-1}$ indicates the cutoff used to select the Splash population. For the Gaussian fit, we found a mean of $\langle V_\phi \rangle$ = 221\,\kms and a standard deviation of $\sigma_{V_{\phi}} = 24$\,\kms for the low-$\alpha$ disk, while for the high-$\alpha$ sample, the values are $\langle V_\phi \rangle$ = 181\,\kms and $\sigma_{V_{\phi}} = 43$\,\kms.} 
    \label{Figure2_nova}
\end{figure}

\begin{figure*}[ht!]
    \centering
    \includegraphics[width=2.1\columnwidth]{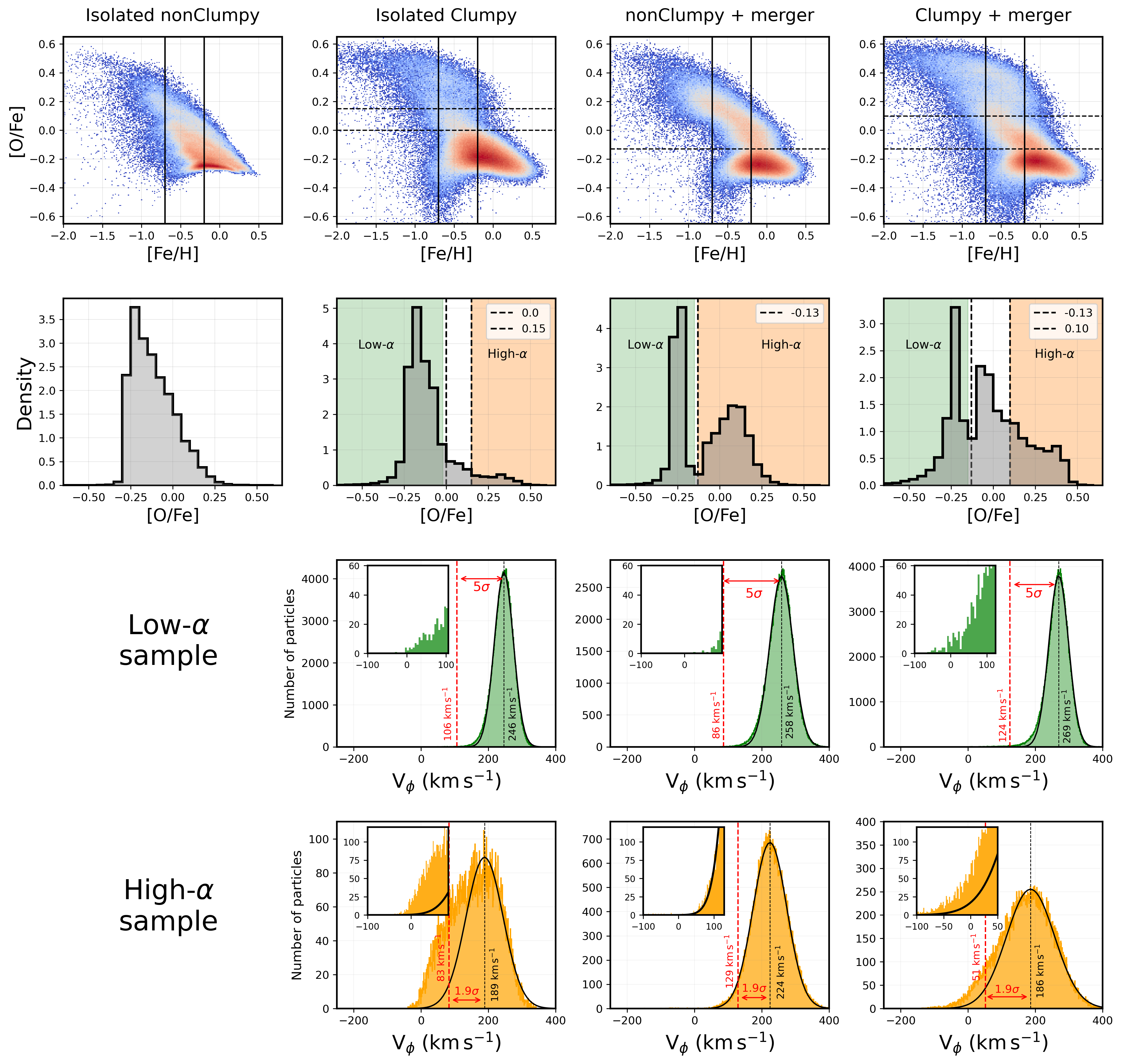}
    \caption{From left to right, the columns show the \texttt{Isolated nonClumpy}, \texttt{Isolated Clumpy}, \texttt{nonClumpy+merger}, and \texttt{Clumpy+merger} simulations. \textbf{First row:} projection onto the [O/Fe]–[Fe/H] plane, where red colors indicate regions of highest stellar density; the vertical lines delimit the interval $-0.7 < \mathrm{[Fe/H]} < -0.2$, and the dashed horizontal line marks the cutoff separating the high- and low-$\alpha$ populations. \textbf{Second row:} [O/Fe] distributions within $-0.7 < \mathrm{[Fe/H]} < -0.2$ for each simulation; the dashed vertical lines indicate the high- and low-$\alpha$ separation. \textbf{Third row:} $V_{\phi}$ distributions for the low-$\alpha$ population in the three simulations that form a bimodality. The inset panels show a zoom of the Splash region, and the dashed vertical line indicates the $\langle V_\phi \rangle$ of each Gaussian fit. \textbf{Fourth row:} same as the third row, but for the high-$\alpha$ population. In the third and fourth rows, the red dashed vertical lines indicate the cutoff used to select the Splash population, accompanied by the exact $V_{\phi}$ adopted.}
    \label{Figure3}
\end{figure*}

Figure \ref{Figure2_nova} shows the $V_{\phi}$ distribution for both low- and high-$\alpha$ populations. A Gaussian fit was applied to each distribution to facilitate visualization of the canonical disk component. To define the Splash population, we select stars with halo-like azimuthal velocities, $V_{\phi} < 100$\,\kms, following the selection criterion adopted by \citet{Belokurov2020}. Note that this cutoff corresponds to distances of 5$\sigma$ and 1.9$\sigma$ from the medians of the Gaussian distributions for the low-$\alpha$ and high-$\alpha$ disks, respectively. We find that Splash stars correspond to the tail of the canonical disk distributions, i.e., the excess of stars beyond the Gaussian distribution. For the high-$\alpha$ sample, however, our adopted cutoff appears to include a fraction of stars that still belong to the Gaussian component, likely associated with the low-velocity tail of the canonical high-$\alpha$ disk. \citet{Amarante2020} found that approximately 13\% of thick disk stars exhibit halo-like kinematics and can therefore be misclassified as members of the Splash population. This explains the visual difference in the $0 < V_{\phi}\,(\kms) < 100$ region between the inner panels of Figure~\ref{Figure2_nova}. 

In this work, all stars below the $V_{\phi} < 100$\,\kms are classified as Splash, yielding 169 low-$\alpha$ and 2095 high-$\alpha$ Splash stars. These represent a small fraction of each disk population (Figure~\ref{Figure2_nova}, bottom). It is worth noting that stars with thin disk chemistry and thick disk kinematics ($100 \lesssim V_{\phi} (\kms) \lesssim 130$) may also be present and could have been affected by the same mechanism that gave rise to the Splash population. Although such stars can be identified in our APOGEE sample, we do not include them in our analysis to preserve a cleaner association between the Splash and a halo-like population.

Figure~\ref{Figure1_nova} shows the chemical and dynamical distributions of low-$\alpha$ (green) and high-$\alpha$ (orange) Splash stars. Since the low-$\alpha$ region at $[\mathrm{Fe/H}] < -0.6$ may include contamination from stars accreted from dwarf galaxies, such as the GSE (e.g., \citealt{Feuillet2020, Naidu2021, Limberg2022}), we highlight the low-$\alpha$ Splash stars with $[\mathrm{Fe/H}] > -0.6$, which are expected to be mainly of in-situ origin, as the known major merger does not extend to such high metallicities. 
Gray contours indicate the canonical low-$\alpha$ disk for comparison. In the Toomre diagram ($\sqrt{V_R^2 + V_Z^2}$ versus $V_{\phi}$, top-right panel in Figure \ref{Figure1_nova}), the low-$\alpha$ Splash is distributed across the halo region, distinct from the canonical thin disk ($V_{\phi} \geq 200$\,\kms). Despite exhibiting chemical compositions characteristic of the thin disk, these stars display higher orbital eccentricities and extend to vertical heights ($|Z_{\mathrm{GC}}|$) significantly above the thin disk scale height of $h_z = 0.3$\,kpc \citep{Juric2008}. Note that almost all the stars in both parts of the Splash have $e > 0.6$, which is consistent with \citet{Kisku2025} definition for Splash selection based on an eccentricity cutoff. 

The kinematic similarity between the low- and high-$\alpha$ Splash, that is, halo-like orbits with low-$V_\phi$, high eccentricity, and that reach great heights relative to the plane, suggests a common dynamical origin. The high-$\alpha$ Splash population is thought to have formed as a consequence of the GSE merger with the pre-existing disk (\citealt{Bonaca2020, Belokurov2020, Lee2023}) or through an early clumpy formation phase at $z \sim 1-2$ \citep{Amarante2020a}. Thus, if the formation of both Splash populations is related, it requires the presence of an ancient low-$\alpha$/thin disk population as described in \citet{Borbolato2025}.


\subsection{Low-$\alpha$ Splash in simulated MW analog galaxy}
\label{subsec:GASTROsplash} 

\citet{Amarante2020a} analysed the formation of the high-$\alpha$ Splash using two isolated models, the \texttt{Isolated nonClumpy} and \texttt{Isolated Clumpy}\footnote{Their clumpy phase is more intense than in our models}. Here, we extend this framework by including two MW analogs undergoing a GSE-like merger, with (\texttt{Clumpy+merger}) and without (\texttt{nonClumpy+merger}) massive star-forming clumps. 

\begin{figure}
    \centering
    \includegraphics[width=\columnwidth]{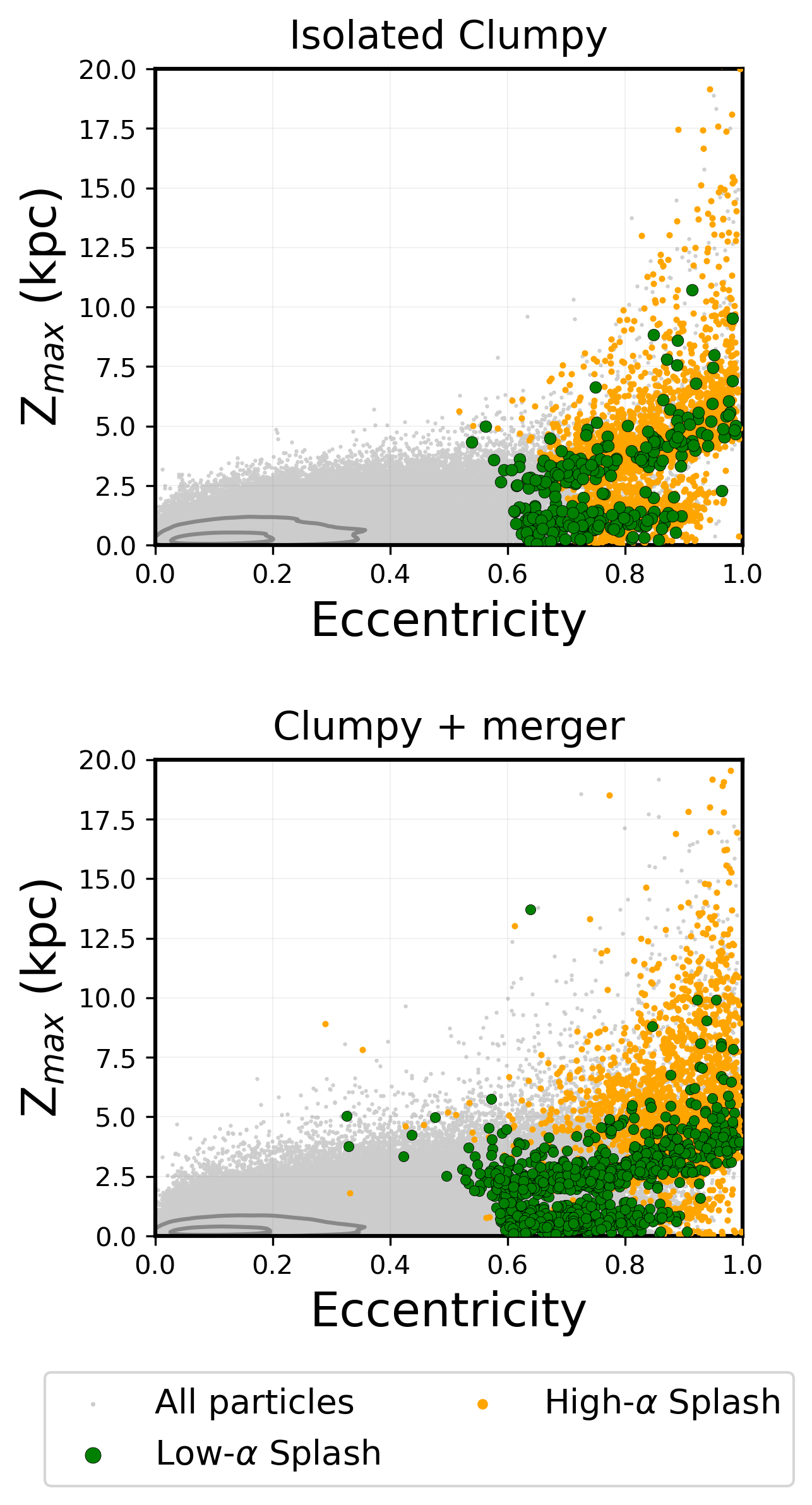}
    \caption{Distribution of high- and low-$\alpha$ Splash populations on the Z$_{\text{max}}$ vs. eccentricity space for the \texttt{Isolated Clumpy} (top) and the \texttt{Clumpy+merger} (bottom panel) models. Gray points represent all particles in the simulation, and the gray contours indicate the higher-density region of the low-$\alpha$/thin disk. The high-$\alpha$ Splash is shown in orange points, while the low-$\alpha$ Splash is shown in green ones.}
    \label{Figure4}
\end{figure}

\begin{figure*}[ht!]
    \centering
    \includegraphics[width=2.1\columnwidth]{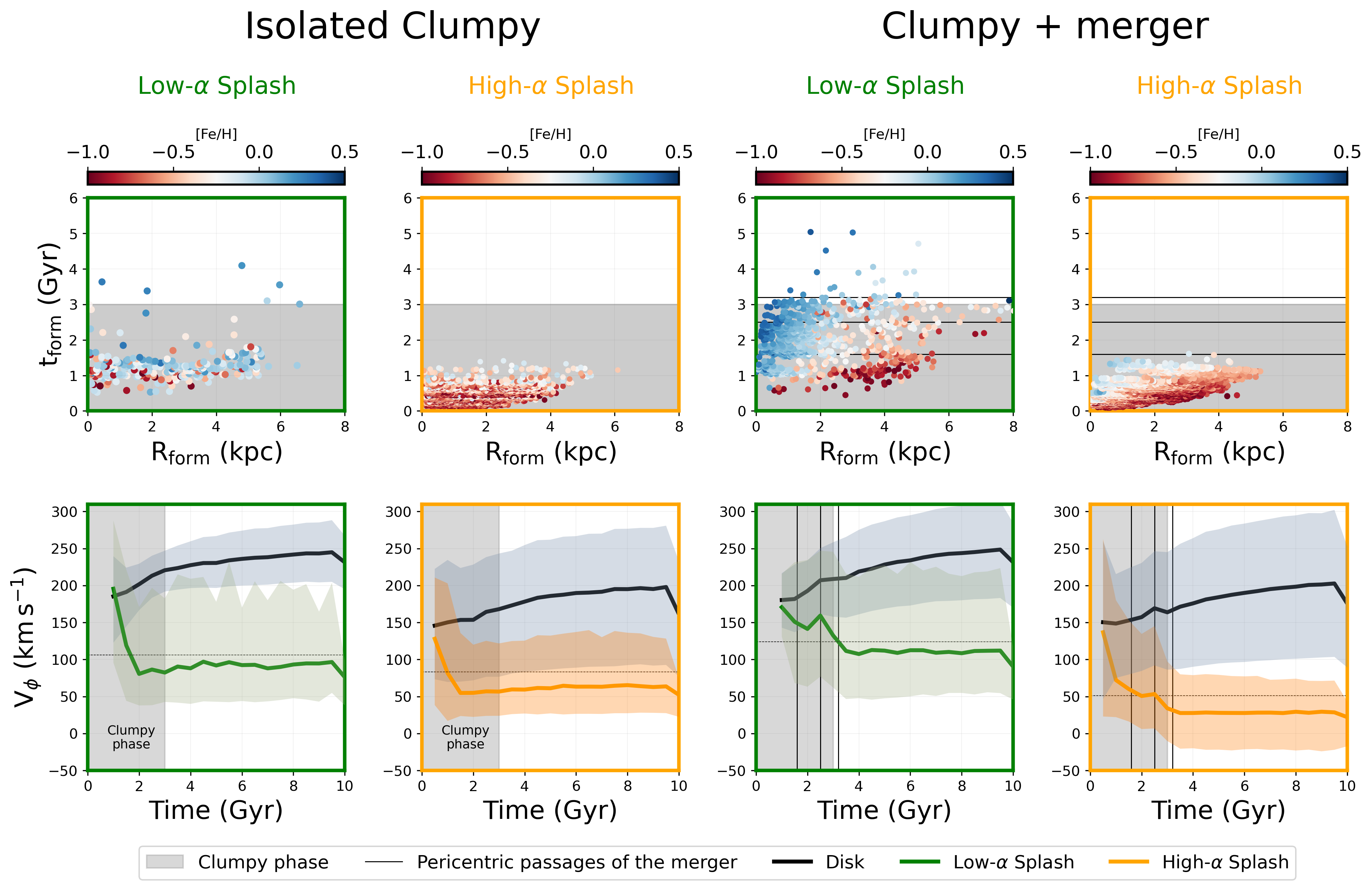}
    \caption{The first two columns show the \texttt{Isolated Clumpy} model, while the last two columns show the \texttt{Clumpy+Merger} model. \textbf{Top row:} distribution of low- and high-$\alpha$ Splash stars in the $\text{t}{_\text{form}}$–$\text{R}{_\text{form}}$ plane, colored by [Fe/H], being the $\text{t}{\text{form}}$ the star formation time and $\text{R}{\text{form}}$ the galactocentric radius at which each star formed. The horizontal lines in the two panels on the right indicate $\text{t}_{\text{form}} = 1.6$, 2.5, and 3.2 Gyr, corresponding to the first, second, and final pericentric passages of the dwarf galaxy, respectively. \textbf{Bottom row:} evolution of mean $V_{\phi}$ as a function of time for the low- (green) and high-$\alpha$ (orange) Splash populations. The Splash populations are represented by the colored lines, while the mean of $V_{\phi}$ for the canonical low- and high-$\alpha$ disks are shown by the upper black line. The shaded area indicates the percentiles of each distribution.The vertical lines mark the same pericentric passages of the dwarf galaxy indicated in the top panels. The gray band in all panels indicates the period of the clumpy phase (first 3\,Gyr).}
    \label{Figure5}
\end{figure*}

Figure~\ref{Figure3} shows the properties of the simulated stellar populations. To better reproduce the same Galactic region used in the observational data, we normalized the disk length by the scale length of each simulation, whose exact values can be found in Table 1 of \citet{amarante+2026}. Only the \texttt{Isolated nonClumpy} model fails to develop the [$\mathrm{O/Fe}$]–[$\mathrm{Fe/H}$]\footnote{In the simulations, Oxygen is the tracer of the $\alpha$-element.} bimodality characteristic of the MW (first row of Figure~\ref{Figure3}). Consequently, we focus on the three bimodal models. As in \citet{amarante+2026}, we separate high- and low-$\alpha$ populations using the $[\mathrm{O/Fe}]$ distribution (second-row panels), constructed solely from stars in the range $-0.7 < [\mathrm{Fe/H}] < -0.2$ (indicated by the vertical lines in the first-row panels). Note that for the clumpy models, we introduced a gap between the two populations to avoid regions where they may overlap. To reproduce the Splash selection criterion adopted for the observational data, we used as a reference the distance between the $V_{\phi}$ cutoff and the mean of the fitted Gaussian distribution, expressed in units of the standard deviation ($\sigma$). In the observational data, this distance corresponds to $5\sigma$ for the low-$\alpha$ population and $1.9\sigma$ for the high-$\alpha$ population. We then applied the same criterion to the simulated $V_{\phi}$ distributions, determining the corresponding cutoff value in each simulation to be used as the Splash selection threshold. The resulting cutoff values are shown in the third and fourth rows of Figure~\ref{Figure3} as red dashed vertical lines. 

Being Splash the tail of the Gaussian distribution, only clumpy models produce an extended $V_{\phi}$ tail for both high- and low-$\alpha$ populations. The \texttt{nonClumpy+merger} simulation does not produce a Splash component, contradicting the expectations that a GSE-like merger alone would produce the $\alpha$-rich Splash (see Section~\ref{sec:discussion}). We note that the stellar mass of the GSE-like galaxy in this model is at the lower end of the measured values \citep{lane+2023}. 
We projected the selected Splash populations onto the Z$_{\text{max}}$--Eccentricity space in Figure~\ref{Figure4}, reproducing panel (e) of Figure~\ref{Figure1_nova}. Similarly to the observational data, the $V_{\phi}$ selection criteria in the simulations are effective at isolating stars on halo-like orbits (high eccentricity, $e > 0.6$), i.e. Splash stars. We note that the apparent wedges in the $Z_{\text{max}}$--Eccentricity plane, also seen in the observational data, are a dynamical effect which depends on the potential of the galaxy \citep[][]{Amarante2020, koppelman+2021}.


Figure~\ref{Figure5} explores Splash formation in the \texttt{Isolated Clumpy} and \texttt{Clumpy+merger} models. The top panels show formation radius, $\text{R}_{\text{form}}$, and time of formation, $\text{t}_{\text{form}}$, colored by [Fe/H]. Splash stars in both models form primarily during the first 3\,Gyr, coinciding with the clumpy phase. In the \texttt{Clumpy+merger} case, pericentric passages of the dwarf galaxy are marked at $t = 1.6, 2.5,$ and $3.2$ Gyr. Although stars at R$_{\text{GC}} < 5$ kpc are excluded from the analysis, most Splash stars originate in the inner Galaxy. We also note that, while the Splash population in the \texttt{Isolated Clumpy} model does not exhibit an [Fe/H] gradient with R$_{\text{form}}$, it displays a radial metallicity gradient in the \texttt{Clumpy+Merger} model. The formation of Splash metallicity gradients in clumpy and non-clumpy MW models merits its own separate investigation. 

The evolution of the mean $V_{\phi}$ (bottom panels of Figure~\ref{Figure5}) shows that, while the disk remains rotationally supported, Splash stars are rapidly heated in all the clumpy models, with or without mergers, indicating that clumps dominate the process. For the \texttt{Clumpy+merger} model, we observe a peak in the mean $V_{\phi}$ immediately following the second pericentric passage of the dwarf. The decline in $V_{\phi}$ at 10\,Gyr, in both models, is a selection artifact due to the criteria used to define Splash stars. We verify that this drop in median $V_{\phi}$ also occurs when the selection is made at previous snapshots.

\section{Discussion}
\label{sec:discussion}

\citet{Borbolato2025} presented observational evidence for the co-formation of the thick/high-$\alpha$ and thin/low-$\alpha$ disks using precise stellar ages from the isochrone-fitting \texttt{StarHorse} code \citep{Queiroz2023}. In particular, they found that the thin disk was already in place before the GSE dwarf merger, reinforcing previous studies indicating an early thin disk formation (e.g., \citealt{Aguirre2018, Beraldo2021, Nepal2024, Gent2024}). Therefore, the presence of a low-$\alpha$ Splash component may also indicate the presence of an old thin disk, independent of age estimates. The idea that the Splash population may also include low-$\alpha$ stars was shown in \citet{Nepal2024}, based on observational data from Gaia DR3 Radial Velocity Spectrometer (RVS) spectra \citep{Cropper2018GaiaDR2rvs}, chemical abundances ([$\alpha$/Fe] and [Fe/H]) derived using a hybrid convolutional neural network method \citep{Guiglion2024} and ages from the \texttt{StarHorse} code. They showed that the region traditionally associated with the Splash in the $V_{\phi}$--[Fe/H] diagram is also populated by low-$\alpha$ stars older than 9\,Gyr. However, they did not investigate the mechanisms responsible for the formation of this population and its clear association with the high-$\alpha$ Splash. Some low-$\alpha$ Splash stars are also visible in \citet{Kisku2025} (see their Figure 6), although they are not discussed. In addition, we verified that the Eos population (\citealt{Myeong2022, Davies2025}) corresponds to the most metal-poor part ([Fe/H] $<$ -0.3) of the low-$\alpha$ Splash.

We suggest a common dynamical heating mechanism between the low- and high-$\alpha$ Splash related to the presence of massive star-forming clumps in the early times, given that the Splash in our models appears only in clumpy MW-like galaxies. A possible explanation is that the dynamic heating of these stars may occur via scattering by clumps, a mechanism already studied in previous works (e.g., \citealt{Elmegreen2016, Struck2026}) and can affect both chemical populations. Also, clumpy models are uniquely able to reproduce both the Splash population and the old (Age $>$ 10\,Gyr) low-$\alpha$ disk. Our results provide a counterpoint to recent literature on the origin of low-$\alpha$ metal-rich halo stars. While \citet{Lee2023} attributed this population up to [$\mathrm{Fe/H}$] $<$ -0.6 to GSE accretion, this does not explain the low-$\alpha$ Splash stars we observe at [$\mathrm{Fe/H}] > -0.6$. While the low-metallicity regime may include contamination from dwarf galaxies such as the GSE (e.g., \citealt{Feuillet2020, Naidu2021, Limberg2022}), no known major merger reaches [Fe/H]$ > -0.6$, implying an in-situ origin. Furthermore, the [Fe/H] range of the low-$\alpha$ Splash is consistent with that of the old low-$\alpha$ disk (e.g., \citealt{Aguirre2018, Nepal2024, Borbolato2025}). Note that the low-$\alpha$ stars of solar and supersolar metallicity dating from the first billion years in the \texttt{Clumpy+merger} model are born in the central region of the Galaxy, where faster chemical enrichment is expected since early stages (see Figure \ref{Figure5}). 

Alternative formation scenarios for the Splash have been proposed, including minor mergers or bar resonances. While some models propose that multiple retrograde minor mergers could form the Splash (e.g. \citealt{Kisku2025}), the observed preference for prograde minor mergers in the MW makes this unlikely \citep{Malhan2022}. Likewise, the Galactic bar is unlikely to explain retrograde stars in the solar neighborhood. \citet{Fiteni2021} showed that bar-driven retrograde stars remain near the bar vicinity, while retrograde stars in the solar neighborhood, in clumpy models, are exclusively clump-driven. Finally, we note a discrepancy with some cosmological simulations that favor a major-merger origin for the Splash \citep[e.g][]{Grand2020,Liao2024, orkney+2025}. We argue that although the GASTRO simulations are more tailored-made experiments, with a controlled merger configuration that reproduces the GSE dwarf merger \citep{Amarante2022}, it still provides an accurate representation of the formation history of the MW's disk \citep{amarante+2026}.


Our findings align with \citet{Amarante2020a}, who demonstrated that the high-$\alpha$ Splash arises from clump scattering in the early disk, and with \citet{Buder2025}, who link the Splash to the dispersion-dominated nature of the proto-galaxy. In the latter case, extragalactic studies similarly showed that massive clumps strongly affect velocity fields and enhance the disk velocity dispersion (e.g., \citealt{Erb2004, Weiner2006, Dekel2009}), contributing to the dispersion-dominated nature of early disks.



\section{Conclusions}
\label{conclusion} 

This Letter has investigated the existence of a low-$\alpha$ Splash in the MW. For this goal, we use spectroscopic data of red giant stars from APOGEE DR17 with distances from \texttt{StarHorse} and astrometry from \textit{Gaia} DR3. Furthermore, we investigated the Splash formation (low- and high-$\alpha$) using GASTRO suite simulation focused on reproducing the main characteristics of the MW. We consider two main scenarios for the formation of such a population: massive star-forming clumps in the early disk or the MW’s last major merger with the GSE disrupted dwarf galaxy. Our main conclusions are summarized as follows:

\begin{itemize}

    \item [(i)] We report the existence of stars chemically compatible with the low-$\alpha$/thin disk on halo-like orbits in the APOGEE sample of red giant stars, and dub this population as low-$\alpha$ Splash. 

    \item [(ii)] We argue that the low-$\alpha$ Splash was formed by the same mechanism that gave rise to the traditional high-$\alpha$ Splash, implying that an old low-$\alpha$/thin disk must have already existed since the first billion years of the MW. This ancient low-$\alpha$ disk population has been discussed in previous works with stellar ages. However, the low-$\alpha$ Splash may constitute observational evidence that is independent of these age estimates.

    \item [(iii)] We show that the high- and low-$\alpha$ Splash are formed only in the clumpy MW models, possibly formed by the scattering of clumps, without the need for the merger. This result is compatible with the proposals of \citet{Amarante2020a} and \citet{Buder2025}, and also aligns with the need for the presence of clumps during the formation of the proto-disk, required to produce old low-$\alpha$ stars by the clumpy model.

    \item [(iv)] A GSE-like merger does not perturb disk stars sufficiently for them to acquire the halo-like kinematics that define the Splash population.
\end{itemize}

The Splash population, now separated into high-$\alpha$ and low-$\alpha$ components, remains important in investigating the early phases of the MW, as it is associated with the Galactic proto-disk. Detailed analysis of these populations can reveal important insights into the mechanisms that governed the disk’s evolution at that time.






\acknowledgements

L.B. thanks the partial financial support by the São Paulo Research Foundation (FAPESP), Brasil (Proc. 2024/16510-2), CAPES/PROEX (proc. 88887.821814/2023-00), and also thanks to all those involved with the multi-institutional \textit{Milky Way BR} Group for the weekly discussions.
JA and ZYL are supported by the National Natural Science Foundation of China under grant No. 12233001, by the National Key R\&D Program of China under grant No. 2024YFA1611602, by a Shanghai Natural Science Research Grant (24ZR1491200), by the ``111'' project of the Ministry of Education under grant No. B20019, and by the China Manned Space Project with No. CMS-CSST-2021-B03. JA and ZYL also thank the sponsorship from Yangyang Development Fund. 
S.R. also thanks partial financial support from FAPESP (Proc.  2020/15245-2), CNPq (Proc. 303816/2022-8), and CAPES.
TK acknowledges support from the NSFC (Grant No. 12303013) and support from the China Postdoctoral Science Foundation (Grant No. 2023M732250).

This work has made use of data from the European Space Agency (ESA) mission {\it Gaia} (\url{https://www.cosmos.esa.int/gaia}), processed by the {\it Gaia} Data Processing and Analysis Consortium (DPAC, \url{https://www.cosmos.esa.int/web/gaia/dpac/consortium}). Funding for the DPAC has been provided by national institutions, in particular, the institutions participating in the {\it Gaia} Multilateral Agreement.

\clearpage

\bibliographystyle{aasjournal}

\bibliography{Bibliography.bib}


\end{document}